\documentclass[aip,jcp,reprint,floatfix]{revtex4-1}

\usepackage[colorlinks=true,citecolor={blue},urlcolor={blue}, linkcolor=blue]{hyperref}%

\usepackage{bm}
\usepackage[normalem]{ulem}
\usepackage[dvipdfmx]{graphicx}
\usepackage{dcolumn}
\usepackage{color}
\usepackage{amsmath}
\usepackage{amssymb}
\usepackage{amsfonts}

\draft 

\begin{document}


\title{Time-dependent optimized coupled-cluster method for multielectron
dynamics II. A coupled electron-pair approximation
}



\author{Himadri Pathak}
\email[Electronic mail:]{pathak@atto.t.u-tokyo.ac.jp}
\affiliation{
Department of Nuclear Engineering and Management, School of Engineering,
The University of Tokyo, 7-3-1 Hongo, Bunkyo-ku, Tokyo 113-8656, Japan
}
\author{Takeshi Sato}
\email[Electronic mail:]{sato@atto.t.u-tokyo.ac.jp}
\affiliation{
Department of Nuclear Engineering and Management, School of Engineering,
The University of Tokyo, 7-3-1 Hongo, Bunkyo-ku, Tokyo 113-8656, Japan
}
\affiliation{
Photon Science Center, School of Engineering, 
The University of Tokyo, 7-3-1 Hongo, Bunkyo-ku, Tokyo 113-8656, Japan
}
\affiliation{
Research Institute for Photon Science and Laser Technology, 
The University of Tokyo, 7-3-1 Hongo, Bunkyo-ku, Tokyo 113-0033, Japan
}
\author{Kenichi L. Ishikawa}
\email[Electronic mail:]{ishiken@n.t.u-tokyo.ac.jp}
\affiliation{
Department of Nuclear Engineering and Management, School of Engineering,
The University of Tokyo, 7-3-1 Hongo, Bunkyo-ku, Tokyo 113-8656, Japan
}
\affiliation{
Photon Science Center, School of Engineering, 
The University of Tokyo, 7-3-1 Hongo, Bunkyo-ku, Tokyo 113-8656, Japan
}
\affiliation{
Research Institute for Photon Science and Laser Technology, 
The University of Tokyo, 7-3-1 Hongo, Bunkyo-ku, Tokyo 113-0033, Japan
}


\date{\today}

\begin{abstract}
We report the implementation of a cost-effective approximation method
within the framework of time-dependent optimized coupled-cluster (TD-OCC) method
[J. Chem. Phys. {\bf 148}, 051101 (2018)] for real-time simulations of
intense laser-driven multielectron dynamics. The method, designated as TD-OCEPA0, is a time-dependent
extension of the simplest version of the coupled-electron pair
approximation with optimized orbitals [J. Chem. Phys. {\bf 139}, 054104 (2013)]. It is size extensive, gauge invariant,
and computationally much more efficient than the TD-OCC with
double excitations (TD-OCCD). We employed this method to simulate the
electron dynamics in Ne and Ar atoms exposed to intense near infrared
laser pulses with various intensities.
The computed results, including high-harmonic generation spectra and
 ionization yields, are compared with those of various other methods
 ranging from uncorrelated time-dependent Hartree-Fock (TDHF) to fully-correlated
 (within the active orbital space) time-dependent complete-active-space
 self-consistent-field (TD-CASSCF). 
The TD-OCEPA0 results show a good agreement with TD-CASSCF ones for 
moderate laser intensities. For higher intensities, however, TD-OCEPA0
tends to overestimate the correlation effect, as occasionally observed for CEPA0 in the ground-state correlation
energy calculations. 
\end{abstract}


\maketitle 


\section{introduction\label{sec:sec1}}
In recent years there has been a significant breakthrough in the experimental techniques to measure and control the motions
of electrons in atoms and molecules, for example, measurement of the delay in photoionization \cite{schultze2010, klunder2011probing}, migration
of charge in a chemical process \cite{belshaw2012observation, calegari2014}
and dynamical change of the orbital picture during the course of bond breaking or formation \cite{itatani2004, smirnova2009high, haessler2010attosecond}.
Atoms and molecules interacting with laser pulses of intensity
$10^{14}\,{\text{W}}/{\text {cm}}^2$ or higher in the visible to
mid-infrared region, show {\color{black}highly} nonlinear response to the fields such as above-threshold
ionization (ATI), tunneling ionization, nonsequential double ionization (NSDI) and high-harmonic generation (HHG).
All these phenomena are by nature nonperturbative \cite{protopapas1997atomic}.
The essence of the attosecond science lies in the HHG \cite{ivanovrev2009, agostini2004physics, gallmann2012attosecond}, one of the most  
successful means to generate ultrashort coherent light pulses in the wavelength ranging from extreme-ultraviolet (XUV)
to the soft x-ray regions \cite{chang2016fundamentals, zhao2012tailoring, takahashi2013attosecond, popmintchev2012bright},
which can be used to unravel the electronic structure \cite{itatani2004, smirnova2009high} or dynamics \cite{calegari2014, kraus2015measurement} of many-body quantum systems.
The HHG spectrum is characterized by a plateau where the
intensity of the emitted light remains nearly constant up to many orders, followed by a sharp cutoff \cite{kli2010}.

A whole lot of numerical methods have been developed to understand atomic and molecular dynamics in the intense
laser field (For a comprehensive review on various wavefunction-based methods for the study of laser-induced electron dynamics,  see Ref.~\citenum{ishikawa2015review})
to catch up with the progress in the experimental techniques.
In principle, ``the best'' one could do is to solve the time-dependent Schr{\"o}dinger equation (TDSE) to have an exact description. 
However, the exact solution of the TDSE is not feasible for systems
containing more than two electrons
\cite{parker1998intense, parker2000time, pindzola1998time, laulan2003correlation, ishikawa2005above, feist2009probing, ishikawa2012competition, sukiasyan2012attosecond, vanroose2006double, horner2008classical}.
As a consequence, single-active electron (SAE) approximation has been
widely used, in which 
the outermost electron is explicitly treated with the effect
of the other electrons modeled by an effective potential.
The SAE model has been successful in numerically {\color{black}exploring} various
high-field phenomena \cite{krause1992jl,
kulander1987time}.
{\color{black}However, the missing electron correlation in SAE makes this method at best qualitative. \cite{haessler2010attosecond}}

Among other established methods, the multiconfiguration time-dependent Hartree-Fock (MCTDHF)
\cite{caillat2005correlated, kato2004time, nest2005multiconfiguration, haxton2011multiconfiguration, hochstuhl2011two} and
time-dependent complete-active-space self-consistent-field (TD-CASSCF) \cite{sato2013time, sato2016time, tikhomirov2017high} are
the most competent theoretical methods for the study of the laser-driven multielectron dynamics
where both configuration interaction (CI) coefficients and the orbitals are propagated in time.
The time-dependent (or {\it optimized} in the sense of time-dependent
variational principle) orbital formulation widens the
applicability of these methods by allowing to use a fewer number of
orbitals than the case of fixed orbital treatments.
{\color{black}Though powerful}, the dilemma with these full CI-based methods is the applicability to large atomic or molecular systems
due to the factorial escalation of the computational cost with respect to the number of electrons. 
To subjugate this difficulty, more approximate, thus computationally
more efficient time-dependent multiconfiguration self-consistent-field (TD-MCSCF)
methods have been developed, based on the truncated CI expansion within the chosen active
orbital space \cite{miyagi2013time, miyagi2014time, haxton2015two,sato2015time}
{\color{black}, compromising size extensivity.}

{\color{black}To} regain the size extensivity, the coupled-cluster
expansion \cite{shavitt:2009, kummel:2003, crawford2007introduction} of
the time-dependent wavefunction emerges naturally as an alternative to
the truncated CI expansion.
The initial ideas of developing time-dependent coupled-cluster
go back to as early as 1978 by Sch{\"o}nhammer and Gunnarsson
\cite{schonhammer1978schonhammer}, and Hoodbhoy and Negele
\cite{hoodbhoy1978time, hoodbhoy1979time}.
Here we take note of a few theoretical works on the time-dependent
coupled-cluster method for time-independent Hamiltonian\cite{dalgaard1983some, koch1990coupled, takahashi1986time, prasad1988time, sebastian1985correlation} 
and a few recent applications of the method with fixed orbitals\cite{pigg2012time, nascimento2016linear}.
Huber and Klamroth \cite{huber2011explicitly} were the first to apply
the coupled-cluster method (with single and double excitations) to
laser-driven dynamics of molecules, using time-independent orbitals and 
the CI wavefunction reconstructed from the propagated CC amplitudes to
evaluate {\color{black}the expectation value of operators}.

In {\color{black}2012}, Kvaal pioneered a time-dependent coupled-cluster method using
time-dependent orbitals for electron dynamics, designated as orbital-adaptive time-dependent
coupled-cluster (OATDCC) method \cite{kvaal2012ab}.
Based on Arponen's bi-orthogonal formulation of the coupled-cluster theory\cite{arponen1983variational}, 
the OATDCC method is derived from the complex-analytic action functional
using time-dependent biorthonormal orbitals. 
Recently, we have also developed 
time-dependent optimized coupled-cluster (TD-OCC) method
\cite{sato2018communication} based on the
real action functional using time-dependent orthonormal orbitals.
The TD-OCC method is a time-dependent extension of the orbital optimized
coupled-cluster method popular in the stationary electronic structure theory
\cite{scuseria1987optimization, sherrill1998energies, krylov1998size,
lindh2012optimized}.
It is not only size extensive, but also gauge invariant, and scales
polynomially with respect to the number of active electrons $N$.
Theoretical as well as numerical comparison of closely-related OATDCC
and TD-OCC methods is yet to be done, and will be discussed elsewhere. 
(See Refs.~\citenum{Myhre:2018} and \citenum{Pedersen:2019} for recent
theoretical accounts on orbital-optimized and time-dependent
coupled-cluster methods, and Refs.~\citenum{pedersen1999gauge} and \citenum{pedersen2001gauge2} 
for the gauge-invariant coupled-cluster response theory with orthonormal
and biorthonormal orbitals, respectively.)

We have implemented TD-OCC method with double excitations (TD-OCCD) and
double and triple excitations (TD-OCCDT) within the chosen active
space\cite{sato2018communication}, of which the computational cost
scales as $N^6$ and $N^8$, respectively. Such scalings are milder than the
factorial one in the MCTDHF and TD-CASSCF methods; {\color{black}nevertheless}, a lower cost
alternative within the TD-OCC framework is highly appreciated to
further extend the applicability to heavier atoms and larger
molecules interacting with intense laser fields.

One such low-cost model is a family of the methods called coupled-electron pair approximation
(CEPA), originally introduced in 1970's \cite{meyer1971ionization, meyer1977methods,
werner1976pno, ahlrichs1985coupled, pulay1985variational, ahlrichs1975pno, koch1980comparison}. 
In particular, the simplest version of the family, denoted as CEPA0 (See Sec.~\ref{sec2}
for the definition.), is recently attracting a renewed attention
\cite{wennmohs2008comparative, neese2009efficient,kollmar2010coupled,malrieu2010ability} 
due to its high {\color{black}cost-performance} balance. The orbital-optimized version of
this method (OCEPA0) has been also developed and applied to the
calculation of, e.g, equilibrium geometries and harmonic vibrational
frequencies of molecules
\cite{bozkaya2013orbital}, which motivated us to extend it to the time-dependent problem.

{\color{black}In} the present article, we report the implementation of the time-dependent, orbital-optimized version of
the CEPA0 theory, hereafter referred to as TD-OCEPA0.
Pilot applications to the simulation of induced dipole moment,
high-harmonic spectra, and ionization probability in three different laser intensities for Ne and Ar are reported. We compare
TD-OCEPA0 results with those of other methods ranging from
uncorrelated TDHF, 
{\color{black}TD-MCSCF with a truncated CI expansion},
TD-OCCD, and fully correlated TD-CASSCF, using the 
same number of active orbitals (except for TDHF) to quantitatively explore the
performance of TD-OCEPA0.
The computational cost of TD-OCEPA0 scales as $N^6$, which is formally
the same as that of TD-OCCD; however as shown in Sec.~\ref{sec2}, 
one need not solve for the double deexcitation operator $\Lambda_2$, but it is sufficient to
propagate the double excitation operator $T_2$ since $\Lambda_2$=$T_2^\dag$.
This leads to a great saving of the computational time
as numerically demonstrated in Sec.~\ref{sec3}.


The manuscript is arranged as follows. A concise description of the TD-OCEPA0 method is presented in Sec. \ref{sec2}.
{\color{black}Section}~\ref{sec3} {\color{black}discusses} the {\color{black}computational} results.
Finally, we made our concluding remark in Sec. \ref{sec4}. We {\color{black}use} Hartree atomic units unless stated otherwise, and
Einstein convention is implied throughout for summation over orbital indices.

\section{Method\label{sec2}}
\subsection{Background}\label{sec2-1}
We consider a system with $N$ electrons governed by the following Hamiltonian,
\begin{eqnarray}\label{eq:ham1q}
H &=& \sum_{i=1}^N h(\bm{r}_i,\bm{p}_i) + \sum_{i=1}^{N-1}\sum_{j=2}^N \frac{1}{|\bm{r}_i-\bm{r}_j|},
\end{eqnarray}
where $\bm{r}_i$ and $\bm{p}_i$ are the position and canonical momentum
of an electron {\color{black}$i$}. The corresponding second quantized Hamiltonian reads
\begin{eqnarray}\label{eq:ham2q}
\hat{H}
&=& h^\mu_\nu \hat{E}^\mu_\nu +
 \frac{1}{2}u^{\mu\gamma}_{\nu\lambda} \hat{E}^{\mu\gamma}_{\nu\lambda},
\end{eqnarray}
where $\hat{E}^\mu_\nu = \hat{c}^\dagger_\mu\hat{c}_\nu$ and 
$\hat{E}^{\mu\gamma}_{\nu\lambda} =
\hat{c}^\dagger_\mu\hat{c}^\dagger_\gamma\hat{c}_\lambda\hat{c}_\nu$,
with $\hat{c}^\dagger_\mu$ ($\hat{c}_\mu$) being a creation
(annihilation) operator in a complete,
orthonormal set
{\color{black} of $2n_{\rm bas}$ spin-orbitals $\{\psi_\mu\}$,
where $n_{\rm bas}$ is the number of basis functions (or the number
of grid points) to expand the spatial part of $\psi_\mu$}, 
and
\begin{eqnarray}
h^\mu_\nu = \int dx_1 \psi^*_\mu(x_1) h(\bm{r}_1,\bm{p}_1) \psi_\nu(x_1),
\end{eqnarray}
\begin{eqnarray}
 u^{\mu\gamma}_{\nu\lambda} = \int\int dx_1dx_2
 \frac{\psi^*_\mu(x_1)\psi^*_\gamma(x_2)\psi_\nu(x_1)\psi_\lambda(x_2)}{|\bm{r}_1-\bm{r}_2|},
\end{eqnarray}
where $x_i=(\bm{r}_i,\sigma_i)$ is a composite spatial-spin coordinate.

{\color{black}
The complete set of $2n_{\rm bas}$ spin-orbitals (labeled with
$\mu,\nu,\gamma,\lambda$) is divided into $n_{\rm occ}$ {\it occupied} ($o,p,q,r,s$) and
$2n_{\rm bas}-n_{\rm occ}$ {\it virtual} spin-orbitals having
nonzero and vanishing occupations, respectively, in the coupled-cluster
(or MCSCF) expansion of the total wavefunction. The occupied spin-orbitals
are classified into $n_{\rm core}$ {\it core} spin-orbitals 
which are occupied in the reference $\Phi$ and kept uncorrelated, and
$n_{\rm act}=n_{\rm occ}-n_{\rm core}$ {\it active} spin-orbitals ($t,u,v,w$) among which the
$N_{\rm act}=N-n_{\rm core}$ active electrons are correlated. The active
spin-orbitals are further splitted into those in the {\it hole} space
($i,j,k,l$) and the {\it particle} space ($a,b,c,d$), which are occupied and
unoccupied, respectively, in $\Phi$.
The core spin-orbitals can also be splitted into those in
the {\it frozen-core} space ($i^{\prime\prime},j^{\prime\prime}$) which are fixed in time, and
the {\it dynamical-core} space ($i^\prime,j^\prime$) which are propagated in time\cite{sato2013time}. 

Hereafter we refer to spin-orbitals simply as orbitals.
[Note that Refs.~\citenum{sato2013time,sato2015time,sato2016time} for
TD-MCSCF methods deal with the equations of motion (EOMs) for spatial orbitals.]
}
The system Hamiltonian (\ref{eq:ham2q}) is equivalently written as
\begin{eqnarray}
\hat{H}
&=& E_0 + f^\mu_\nu \{\hat{E}^\mu_\nu\} +
 \frac{1}{4}v^{\mu\gamma}_{\nu\lambda} \{\hat{E}^{\mu\gamma}_{\nu\lambda}\}, 
\end{eqnarray}
where $E_0 = \langle\Phi|\hat{H}|\Phi\rangle$, $f^\mu_\nu =
h^\mu_\nu + v^{\mu j}_{\nu j}$ ($j$ running over core and hole spaces),
$v^{\mu\gamma}_{\nu\lambda}=u^{\mu\gamma}_{\nu\lambda}-u^{\mu\gamma}_{\lambda\nu}$,  
and the bracket $\{\cdots\}$ implies
that the operator inside is normal ordered relative to the reference. 

\subsection{Review of TD-OCC method}\label{sec2-1}
Let us begin with a generic TD-OCC framework, which
relies on the time-dependent variational principle with
real action functional\cite{sato2018communication},
\begin{eqnarray}
S &=& \label{eq:action}
\operatorname{Re}\int_{t_0}^{t_1} Ldt = \frac{1}{2} \int_{t_0}^{t_1}\left( L + L^*\right) dt,
\end{eqnarray}
\begin{eqnarray}
L &=&\label{eq:Lag}
 \langle\Phi|(1+\hat{\Lambda})e^{-\hat{T}}(\hat{H}-i\frac{\partial}{\partial
 t})e^{\hat{T}}|\Phi\rangle,
\end{eqnarray}
where
\begin{eqnarray}\label{eq:amplitude}
\hat{T}&=&\hat{T}_2+\hat{T}_3\cdots=\tau^{ab}_{ij}\hat{E}^{ab}_{ij}+\tau^{abc}_{ijk}\hat{E}^{abc}_{ijk}\cdots, \\
\hat{\Lambda}&=&\hat{\Lambda}_2+\hat{\Lambda}_3\cdots=\lambda_{ab}^{ij}\hat{E}_{ab}^{ij}+\lambda_{abc}^{ijk}\hat{E}_{abc}^{ijk}\cdots,
\end{eqnarray}
with $\tau^{ab\cdots}_{ij\cdots}$ and $\lambda_{ab\cdots}^{ij\cdots}$
being excitation and deexcitation amplitudes, respectively. 
We require that the action to be stationary, $\delta S=0$, with respect to
the variation of amplitudes $\delta\tau^{ab\cdots}_{ij\cdots}$,
$\delta\lambda_{ab\cdots}^{ij\cdots}$ and orthonormality-conserving
orbital variations $\delta\psi_\mu=\psi_\nu\Delta^\nu_\mu$, with
antiHermitian matrix elements $\Delta^\nu_\mu\equiv\langle\psi_\nu|\delta\psi_\mu\rangle$.

{\color{black}To facilitate formulation of TD-OCEPA0 below,} we transform the Lagrangian into two
equivalent expressions, 
\begin{subequations}\label{eqs:Lag_equiv}
\begin{eqnarray}
L&=&
L_0 +
\langle\Phi|(1+\hat{\Lambda})[\{\hat{H}-i\hat{X}\}e^{\hat{T}}]_c|\Phi\rangle
- i\lambda^{ij\cdots}_{ab\cdots}\dot{\tau}^{ab\cdots}_{ij\cdots},
\nonumber \\ \label{eq:Lag_amp}\\
&=&\label{eq:Lag_dens} 
(h^p_q-iX^p_q)\rho^q_p + \frac{1}{2}u^{pr}_{qs} \rho_{pr}^{qs}
- i\lambda^{ij\cdots}_{ab\cdots}\dot{\tau}^{ab\cdots}_{ij\cdots},
\end{eqnarray}
\end{subequations}
where $\hat{X}=X^\mu_\nu\hat{E}^\mu_\nu$ with
$X^\mu_\nu=\langle\psi_\mu|\dot{\psi}_\nu\rangle$ being antiHermitian,
$L_0 = \langle\Phi|(\hat{H}-i\hat{X})|\Phi\rangle$, and the symbol $[\cdots]_c$
indicates the restriction to diagrammatically connected terms.
The one-electron and two-electron reduced density matrices
(RDMs) $\rho^q_p$ and $\rho^{qs}_{pr}$ are defined, respectively, by 
\begin{eqnarray}
 \rho^q_p&=&\label{eq:1rdm}
\langle\Phi|(1+\hat{\Lambda})e^{-\hat{T}}\hat{E}^p_qe^{\hat{T}}|\Phi\rangle, \\
 \rho^{qs}_{pr}&=&\label{eq:2rdm}
\langle\Phi|(1+\hat{\Lambda})e^{-\hat{T}}\hat{E}^{pr}_{qs}e^{\hat{T}}|\Phi\rangle.
\end{eqnarray}
To benefit later discussions, we separate the one-electron and
two-electron RDMs {\color{black}(1RDM and 2RDM, respectively)} into reference and correlation
contributions,
\begin{eqnarray}
\rho^q_p &=& (\rho_0)^q_p + \gamma^q_p, \\
\rho^{qs}_{pr}&=& (\rho_0)^{qs}_{pr} + \gamma^{qs}_{pr},
\end{eqnarray}
where the reference contributions
$(\rho_0)^q_p = \delta^q_j\delta^j_p$
and $(\rho_0)^{qs}_{pr}= 
\gamma^q_p \delta^s_j \delta^j_r
+\gamma^s_r \delta^q_j\delta^j_p
-\gamma^q_r \delta^s_j\delta^j_p
-\gamma^s_p \delta^q_j\delta^j_r
+\delta^q_j\delta^j_p\delta^s_k\delta^k_r
-\delta^s_j\delta^j_p\delta^q_k\delta^k_r$
($j,k$ running over core and hole spaces)
are independent of the correlation treatment, and the correlation
contributions are defined as
\begin{subequations} \label{eqs:RDMs_corr}
\begin{eqnarray}\label{eqs:1RDM_corr}
 \gamma^q_p&=&\label{eqs:1RDM_corr}
\langle\Phi|(1+\hat{\Lambda})[\{\hat{E}^p_q\}e^{\hat{T}}]_c|\Phi\rangle, \\
 \gamma^{qs}_{pr}&=&\label{eqs:2RDM_corr}
\langle\Phi|(1+\hat{\Lambda})[\{\hat{E}^{pr}_{qs}\}e^{\hat{T}}]_c|\Phi\rangle.
\end{eqnarray}
\end{subequations}
See Ref.~\citenum{sato2018communication} for the derivation of general
TD-OCC EOMs based on the real-valued action
principle outlined here.

\subsection{TD-OCEPA0 method}\label{sec2-2}
Now we define the TD-OCEPA0 method by
(i) including double (de)excitations only
($\hat{T}=\hat{T}_2$, $\hat{\Lambda}=\hat{\Lambda}_2$) and
(ii) linearizing the exponential operator $e^{\hat{T}_2}$,
in the normal-ordered, connected expression of the coupled-cluster
Lagrangian [Eq.~(\ref{eq:Lag_amp})], 
\begin{eqnarray}\label{eq:L_CEPA0}
L &=& L_0 +
 \langle\Phi|(1+\hat{\Lambda}_2)[\{\hat{H}-i\hat{X}\}(1+\hat{T}_2)]_c|\Phi\rangle -
 i\lambda^{ij}_{ab}\dot{\tau}^{ab}_{ij}, \nonumber \\
\end{eqnarray}
and, accordingly the correlation RDMs [Eqs.~(\ref{eqs:RDMs_corr})], 
\begin{subequations} \label{eqs:RDM_CEPA0}
\begin{eqnarray}
 \gamma^q_p&=&\label{eq:1RDM_CEPA0}
\langle\Phi|(1+\hat{\Lambda}_2)[\{\hat{E}^p_q\}(1+\hat{T}_2)]_c|\Phi\rangle, \\
 \gamma^{qs}_{pr}&=&\label{eq:2RDM_CEPA0}
\langle\Phi|(1+\hat{\Lambda}_2)[\{\hat{E}^{pr}_{qs}\}(1+\hat{T}_2)]_c|\Phi\rangle.
\end{eqnarray}
\end{subequations}

Requiring $\delta S/\delta \lambda^{ij}_{ab}(t) = 0$ and $\delta S/\delta \tau^{ab}_{ij}(t)
= 0$, respectively, using $L$ of Eq.~(\ref{eq:L_CEPA0}) derives
\begin{eqnarray}
i\dot{\tau}^{ab}_{ij} &=&
\langle\Phi|\hat{E}^{ij}_{ab}[\{\hat{H}-i\hat{X}\}(1+\hat{T}_2)]_c|\Phi\rangle \nonumber \\
&=&\label{eq:td-ocepa0_t2}
v_{ij}^{ab}-p(ij) \bar{f}_j^k
 \tau_{ik}^{ab}+p(ab)\bar{f}_c^a \tau_{ij}^{cb} \nonumber \\
&+&\frac{1}{2}v_{cd}^{ab}\tau_{ij}^{cd}
+\frac{1}{2} v_{ij}^{kl} \tau_{kl}^{ab}+p(ij)p(ab)
v_{ic}^{ak} \tau_{kj}^{cb},
\end{eqnarray}
\begin{eqnarray}
-i\dot{\lambda}^{ij}_{ab} &=&
\langle\Phi|(1+\hat{\Lambda}_2)[\{\hat{H}-i\hat{X}\}\hat{E}^{ab}_{ij}]_c|\Phi\rangle \nonumber \\
&=&\label{eq:td-ocepa0_l2}
v_{ab}^{ij}-p(ij) \bar{f}_k^i
\lambda_{ab}^{kj}+p(ab)\bar{f}_a^c\lambda_{cb}^{ij} \nonumber \\
&+&\frac{1}{2}v_{ab}^{cd}\lambda_{cd}^{ij}
+\frac{1}{2}v_{kl}^{ij}\lambda_{ab}^{kl}+p(ij)p(ab)
v_{kb}^{cj}\lambda_{ac}^{ik},
\end{eqnarray}
where $\bar{f}^p_q=f^p_q-iX^p_q$, and {\color{black}$p(\mu\nu)$} is the
anti-symmetrizer; ${\color{black}p(\mu\nu)}A_{\mu\nu}=A_{\mu\nu}-A_{\nu\mu}$.
{\color{black}Comparing} Eqs~(\ref{eq:td-ocepa0_t2}) and (\ref{eq:td-ocepa0_l2}),
{\color{black}and noting that the orbitals are orthonormal,}
one sees that the EOM for $\lambda_{ab}^{ij}$ is the complex conjugate of
that for $\tau_{ij}^{ab}$, concluding that
$\hat{\Lambda}_2=\hat{T}_2^\dag$. 
As a result, the first and second terms of Eq.~(\ref{eq:L_CEPA0}) are real, and
\begin{eqnarray}
\operatorname{Im}\int_{t_0}^{t_1}Ldt
&=& \frac{1}{2i}\int_{t_0}^{t_1}(L-L^*)dt \nonumber \\
&=& \frac{1}{2}\left\{\left|\tau^{ab}_{ij}(t_1)\right|^2 -
\left|\tau^{ab}_{ij}(t_0)\right|^2\right\}
\end{eqnarray}
is independent of the integration path and irrelevant in taking its
variation. Therefore, {\it given the orthonormal orbitals}, $L$ is
essentially real, and one could equally base {\color{black}oneself} on
\begin{eqnarray}
S &=& \label{eq:action_ocepa0} \int_{t_0}^{t_1} Ldt, 
\end{eqnarray}
for TD-OCEPA0 ansatz. 
See Sec.~\ref{sec2-3} below for more explicit account of this point. 

Based on the natively real action functional $S$ of
Eq.~(\ref{eq:action_ocepa0}), the equation for 
$X^\mu_\nu=\langle\psi_\mu|\dot{\psi}_\nu\rangle$ is derived by
requiring $\delta S/\delta \Delta^\mu_\nu=0$ using the Lagrangian expression 
of Eq.~(\ref{eq:Lag_dens}) and RDMs of Eqs.~(\ref{eqs:RDM_CEPA0}) to obtain
\begin{eqnarray}
i(X^\nu_p \rho^p_\mu -\rho^\nu_p X^p_\mu)
&=& \label{eq:eomforX_ocepa0}
F^\nu_p \rho^p_\mu - \rho^\nu_pF_p^{\mu *},
\end{eqnarray}
where $F^\mu_p =
\langle\psi_\mu|\hat{F}|\psi_p\rangle$,
\begin{eqnarray}
\hat{F}|\psi_p\rangle &=& \label{eq:gfockoperator}
\hat{h} |\psi_p\rangle +  \hat{W}^r_s|\psi_q\rangle
\rho^{qs}_{or}(\rho^{-1})_p^o, \\
W^r_s(x_1) &=& \label{eq:gfockoperator}
\int dx_2 \frac{\psi^*_r(x_2)\psi_s(x_2)}{|\bm{r}_1-\bm{r}_2|}.
\end{eqnarray}
Subsequent analyses of Eq.~(\ref{eq:eomforX_ocepa0}) are parallel to
those for TD-MCSCF methods\cite{miyagi2013time,sato2015time}, and one
arrives at the orbital EOMs,
{\color{black}
\begin{eqnarray}\label{eq:eom_orb}
i|\dot{\psi_p}\rangle &=&
(\hat{1}-\hat{P})
\hat{F}|\psi_p\rangle + |\psi_q\rangle X^q_p,
\end{eqnarray}
}
where $\hat{1}$ is the identity operator within the orbital space
$\{\psi_\mu\}$, {\color{black}and $\hat{P}=\sum_q|\psi_q\rangle\langle\psi_q|$,} with
non-redundant orbital rotations determined by
{\color{black}
\begin{eqnarray}\label{eq:eom_orb_da}
i(\delta^t_u -\rho^t_u) X^u_{i^\prime}
&=& 
F^t_{i^\prime} - \rho^t_u F_u^{i^\prime\!*},
\end{eqnarray}
}
\begin{eqnarray}\label{eq:eom_orb_p}
i(X^a_j \rho^j_i -\rho^a_b X^b_i)
&=& 
F^a_j \rho^j_i - \rho^a_b F_b^{j*}. 
\end{eqnarray}
{\color{black}
A careful consideration of the frozen-core orbitals within the electric dipole approximation derives
\begin{eqnarray}\label{eq:eom_orb_fa}
iX_\mu^{i^{\prime\prime}}
&=& \left\{
\begin{array}{ll}
 0 & \textrm{(length gauge)}\\
 \pmb{E}(t)\cdot\langle\psi_{i^{\prime\prime}}|\pmb{r}|\psi_\mu\rangle  & \textrm{(velocity gauge)} \\
\end{array}
\right., 
\end{eqnarray}
where $\pmb{E}$ is the external electric field, 
enabling gauge-invariant simulations with frozen-core
orbitals.\cite{sato2016time} 
}

Redundant orbital rotations {\color{black}$\{X^{i^\prime}_{j^\prime}\}$},
$\{X^i_j\}$, and, $\{X^a_b\}$ can be arbitrary
antiHermitian matrix elements. In particular, if one chooses
$X^a_b=X^i_j=0$, the term $-i\hat{X}$ is dropped in Eqs.~(\ref{eq:td-ocepa0_t2}) and (\ref{eq:td-ocepa0_l2}).
Again as a consequence of $\hat{\Lambda}_2=\hat{T}^\dagger_2$,
both 1RDM and 2RDM are Hermitian, of which
the algebraic expression of non-zero elements are given by
\begin{subequations}\label{eqs:td-ocepa0_den}
\begin{eqnarray}
\gamma^{ j }_{ i } &=&\label{eq:td-ocepa0_den1}
- \frac{ 1 }{ 2 }
\lambda^{ k j }_{ c d } 
\tau^{ c d }_{ k i }, 
\gamma^{ b }_{ a } =
\frac{ 1 }{ 2 }
\lambda^{ k l }_{ c a} 
\tau^{ c b }_{ k l }, \\
\gamma^{ c d }_{ a b }&=&
\frac{ 1 }{ 2 }
\lambda^{ k l }_{ a b } 
\tau^{ c d }_{ k l }, 
\gamma^{ k l }_{ i j }=
\frac{ 1 }{ 2 }
\lambda^{ k l }_{ c d } 
\tau^{ c d }_{ i j }, \\
\gamma^{ i a }_{ b j }&=&
\lambda^{ k i }_{ c b }
\tau^{ c a }_{ k j }, 
\gamma^{ i j }_{ a b}
= \lambda_{ a b }^{ i j }, \\
\gamma^{ a b }_{ i j }&=&\label{eq:td-ocepa0_den2}
\tau^{ a b }_{ i j }.
\end{eqnarray}
\end{subequations}
In summary, the TD-OCEPA0 method is defined by the EOMs of $\hat{T}_2$
amplitudes [Eq.~(\ref{eq:td-ocepa0_t2})] and orbitals
[Eq.~(\ref{eq:eom_orb})], with the hole-particle mixing determined by
solving Eq.~(\ref{eq:eomforX_ocepa0}) and RDMs given by
Eqs.~(\ref{eqs:td-ocepa0_den}).
\begin{table}[!b]
\caption{\label{tab:comparison} The ground state total energies of Be and Ne atoms.}
\begin{ruledtabular}
\begin{tabular}{lllcc}
&
Basis &
Method&
\multicolumn{1}{c}{This work\footnotemark[1]} &
\multicolumn{1}{c}{PSI4\cite{psi4}} \\
\hline
\hline
Be & 6-31G$^*$\cite{be_basis}&HF    &$-$14.5667\,\,6405 & $-$14.5667\,\,6403 \\
   &                         &CEPA0  &$-$14.6192\,\,0336 & $-$14.6192\,\,0335\\ 
   &                         &OCEPA0 &$-$14.6196\,\,5019 & $-$14.6196\,\,5018\\
   &                         &OCCD  &$-$14.6138\,\,6552 & \\
   &                         &OCCDT &$-$14.6139\,\,4064 & \\
   &                         &FCI   &$-$14.6139\,\,4255 & $-$14.6135\,\,4253\\
Ne & cc-pVDZ\cite{be_basis}  &HF    &$-$128.4887\,\,7555& $-$128.4887\,\,7555\\
   &                         &CEPA0  &$-$128.6802\,\,1409& $-$128.6802\,\,1409\\
   &                         &OCEPA0 &$-$128.6802\,\,9009& $-$128.6802\,\,9009\\
   &                         &OCCD  &$-$128.6795\,\,9316&                    \\
   &                         &OCCDT &$-$128.6807\,\,2135&                    \\
   &                         &FCI   &$-$128.6808\,\,8113& $-$128.6808\,\,8113\\
\end{tabular}
\end{ruledtabular}
\footnotetext[1]{
The overlap, one-electron, and two-electron repulsion
integrals over Gaussian basis functions are generated using 
Gaussian09 program (Ref.~\citenum{gaussian09}), and used to propagate
EOMs in imaginary time in the orthonormalized Gaussian basis,
with a convergence threshold of 10$^{-15}$ Hartree of energy difference
in subsequent time steps.}
\end{table}

\subsection{Relation to other ansatz}\label{sec2-3}
As the name suggests, the TD-OCEPA0 method is a time-dependent
extension of the stationary OCEPA0 method. In the stationary case, it is
known that CEPA0 \cite{ahlrichs1979many, 
wennmohs2008comparative}, D-MBPT($\infty$) \cite{bartlett1977comparison,
bartlett1979quartic}, third-order expectation value coupled-cluster
[XCC(3)] \cite{bartlett1988expectation}, and the linearized CCD
(LCCD) \cite{bartlett1981many} energy functionals are all equivalent. 
The similar equivalence in the time-dependent case can be
demonstrated by considering the Lagrangian of Eq.~(\ref{eq:Lag}). 
In particular, the XCC Lagrangian can be written as
\begin{eqnarray}
L
&=& \frac{1}{\langle\Phi|e^{\hat{T}^\dagger} e^{\hat{T}}|\Phi\rangle}
\langle\Phi|e^{\hat{T}^\dagger}
(\hat{H}-i\frac{\partial}{\partial t})
e^{\hat{T}}|\Phi\rangle
\nonumber \\
&=& \langle\Phi|\left[e^{\hat{T}^\dagger}
(\hat{H}-i\frac{\partial}{\partial t})
e^{\hat{T}}\right]_{sc}|\Phi\rangle \nonumber \\
&\sim&
\langle\Phi|[(1+\hat{T}^\dagger)
(\hat{H}-i\frac{\partial}{\partial t})
(1+\hat{T})]_{sc}|\Phi\rangle, \nonumber \\
&=&
L_0 + 
\langle\Phi|[(1+\hat{T}^\dagger)
\{\hat{H}-i\hat{X}\}
(1+\hat{T})]_{sc}|\Phi\rangle -i\tau^{ab*}_{ij}\dot{\tau}^{ab}_{ij}, \nonumber \\
\end{eqnarray}
where $[\cdots]_{sc}$ restricts to {\it strongly connected} terms, and
the third line introduces the XCC(3) approximation. This
Lagrangian, which leads to the same working equations as derived in
Sec.~\ref{sec2-2} (with $\hat{\Lambda}_2={\color{black}\hat{T_2}^\dagger}$), emphasizes the Hermitian nature of the TD-OCEPA0,
of which a certain advantage over the standard non-Hermitian
treatment is discussed in Ref.~\citenum{taube2009rethinking} in the stationary case.

The EOMs of the TD-OCEPA0 method is simpler than those of the closely-related TD-OCCD
method (See Appendix~\ref{app:td-occd} for algebraic details of the
TD-OCCD method.) in that all terms quadratic to $\tau^{ab}_{ij}$ are absent in the $\hat{T}_2$ equation
[comparing Eq.~(\ref{eq:td-ocepa0_t2}) and (\ref{eq:td-occd_t2})]
and in the 2RDM expression [comparing Eq.~(\ref{eq:td-ocepa0_den2}) and
(\ref{eq:td-occd_den2})].
It should {\color{black}also be} noted that one need not solve
for
$\hat{\Lambda}_2$ for TD-OCEPA0 since
$\hat{\Lambda}_2=\hat{T}^\dagger_2$, in contrast
{\color{black}to the fact that}
Eq.~(\ref{eq:td-occd_l2}) should be solved for TD-OCCD.
These simplifications make TD-OCEPA0 computationally much more efficient than
TD-OCCD, as numerically demonstrated in Sec.~\ref{sec3}.

\section{Numerical results and discussions\label{sec3}}
\subsection{Ground-state energy\label{sec3-1}}
We have implemented the TD-OCEPA0 method for atom-centered Gaussian
basis functions and spherical finite-element discrete variable representation
(FEDVR) basis for atoms, {\color{black}both with spin-restricted and spin-unrestricted
treatments,} by modifying the TD-OCCD code described in
Ref.~\citenum{sato2018communication}. 
Exploiting the feasibility of the imaginary time relaxation to obtain the
ground state\cite{sato2018communication}, we first computed the
ground-state energy of Be and Ne atoms with standard Gaussian
basis sets, and compare the results with those obtained by PSI4 program
package \cite{psi4}, in which the time-independent OCEPA0 method is
implemented. To facilitate the comparison, the number of
active spatial orbitals {\color{black}$n_{\rm act}/2$} are set to be the same as the number of basis
functions {\color{black}$n_{\rm bas}$}, since this is the only capability of the PSI4
program. In this case, there are no
virtual orbitals (See Sec.~\ref{sec2-1} for the definition.), and
therefore, the first term of Eq.~(\ref{eq:eom_orb}) vanishes.
We also take an option of imaginary-propagating amplitudes only, with all
orbitals frozen at the canonical HF solution, to obtain the
fixed-orbital CEPA0 energy. 
\begin{figure}[b!]
\centering
\includegraphics[width=.8\linewidth]{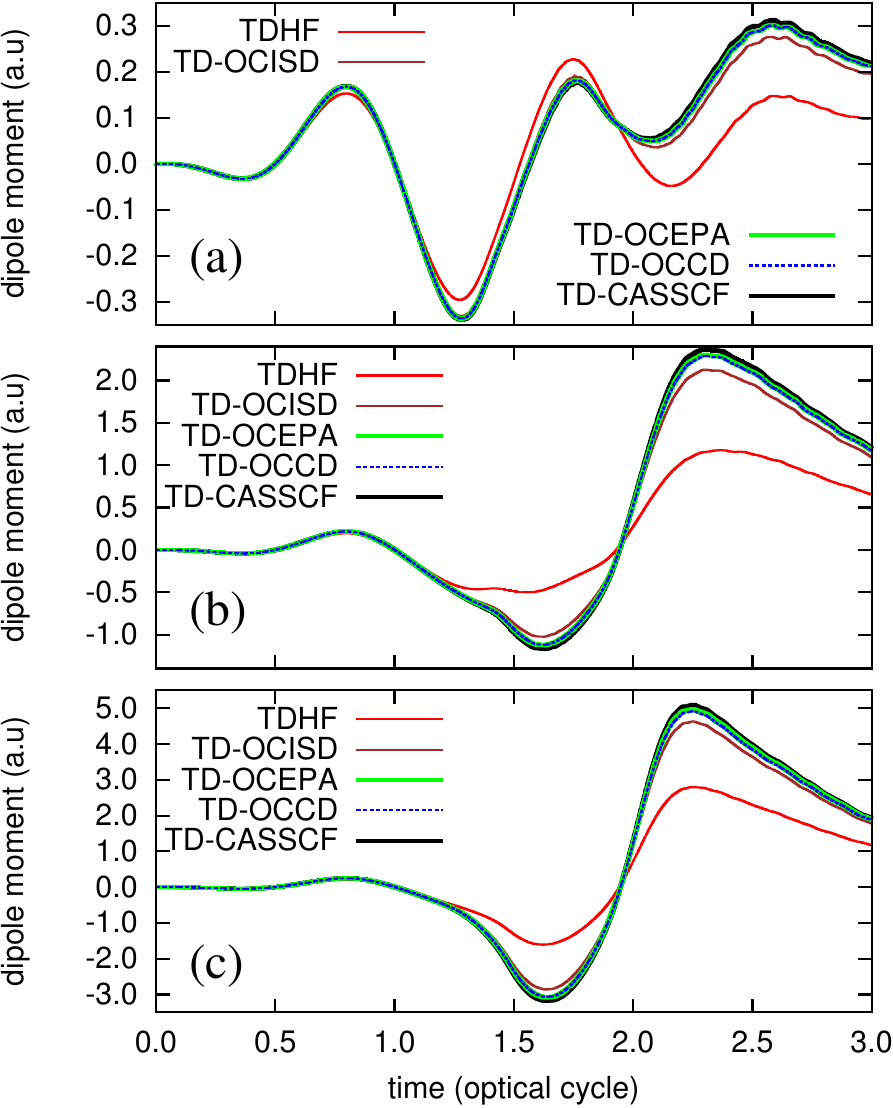}
\caption{\label{fig:nedip}
Time evolution of the dipole moment of Ne irradiated by a laser pulse
of a wavelength of 800 nm and intensities of 5$\times$10$^{14}$ W/cm$^2$
(a), 8$\times$10$^{14}$ W/cm$^2$ (b), and 1$\times$10$^{15}$ W/cm$^2$ (c),
calculated with TDHF, TD-OCISD, TD-OCEPA, TD-OCCD, and TD-CASSCF methods.} 
\end{figure}
\begin{figure}[t!]
\centering
\includegraphics[width=.8\linewidth]{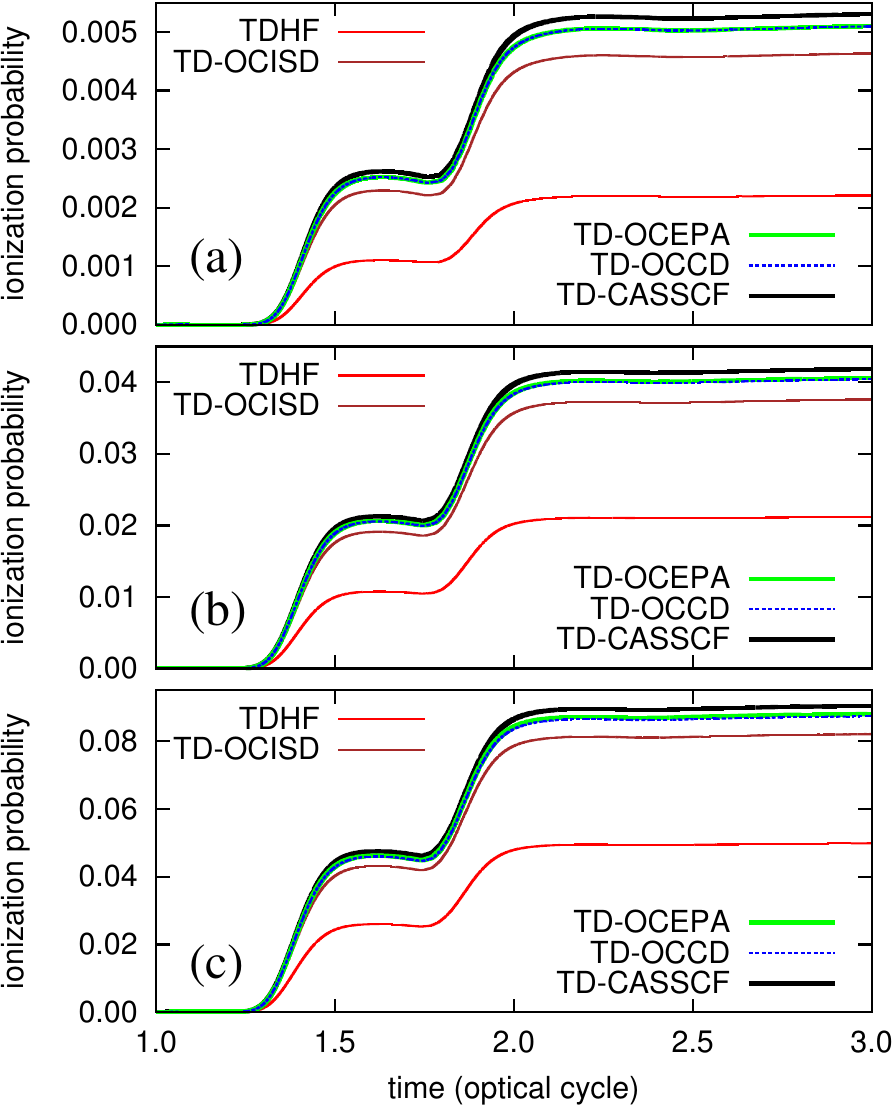}
\caption{\label{fig:neion}
Time evolution of the single ionization
 probability of Ne irradiated by a laser pulse
of a wavelength of 800 nm and intensities of 5$\times$10$^{14}$ W/cm$^2$
(a), 8$\times$10$^{14}$ W/cm$^2$ (b), and 1$\times$10$^{15}$ W/cm$^2$ (c),
calculated with TDHF, TD-OCISD, TD-OCEPA, TD-OCCD, and TD-CASSCF methods.} 
\end{figure}

The calculated total energies listed in Table~\ref{tab:comparison}
demonstrate a virtually perfect agreement of the results of this
work and PSI4 program, both for CEPA0 and OCEPA0 methods and for Be and Ne atoms (with a small discrepancy due to
a digit truncation of Gaussian one- and two-electron integrals), which confirms the correctness of our implementation. 

It is also observed that, for both
Be and Ne, the OCEPA0 energies are noticeably lower than the OCCD ones, and 
for the Be case, the OCEPA0 energy is slightly lower than the FCI energy.
Such an overestimation of the correlation energy is usually
considered not critical, and more than compensated by the size extensivity in the
stationary case. 

\subsection{Application to electron dynamics in Ne and Ar\label{sec3-2}}
Next, we apply the TD-OCEPA0 method to the
laser-driven electron dynamics in Ne and Ar atoms. 
Within the dipole approximation in the velocity gauge, the 
one-electron Hamiltonian is given by
\begin{eqnarray}\label{eq:h1vg}
h(\bm{r},\bm{p})=\frac{1}{2}|\bm{p}|^2 - \frac{Z}{|\bm{r}|} + A(t)p_z, 
\end{eqnarray}
where $Z$ is the atomic number, $A(t)=-\int^t E(t^\prime) dt^\prime$ is
the vector potential, with $E(t)$ being the laser electric field
linearly polarized along $z$ axis. 
It should be noted that TD-OCC method (including TD-OCEPA0) is gauge invariant;
length-gauge and velocity-gauge simulations, upon numerical convergence, give the same
result for observables. The velocity gauge employed here is advantageous
in simulating high-field phenomena\cite{sato2016time,orimo2018implementation}.

The laser electric field is given by
\begin{eqnarray}
E(t)=E_0\,{\text{sin}}(\omega_0t)\,{\text{sin}}^2\left(\pi\frac{t}{3T}\right), 
\end{eqnarray}
for $0 \leq t \leq 3T$, and $E(t)=0$ otherwise, with the central wavelength $\lambda=2\pi/\omega_0=800$ nm, the period
$T=2\pi/\omega_0 \sim 2.67$ fs, and the peak intensity $I_0=E_0^2$. 
We consider three different intensities 5$\times 10^{14}$ W/cm$^2$,
8$\times 10^{14}$ W/cm$^2$, and 1$\times 10^{15}$ W/cm$^2$ for Ne, and 
2$\times 10^{14}$ W/cm$^2$,  4$\times 10^{14}$ W/cm$^2$, and 6$\times 10^{14}$ W/cm$^2$ for Ar.


The orbital functions are represented by the spherical FEDVR basis\cite{sato2016time, orimo2018implementation}, with
the maximum angular momentum $l_{\text{max}}=47$ for Ne and
$l_{\text{max}}=63$ for Ar, and the FEDVR basis supporting the radial
coordinate $0 < r < 240$ using 63 finite elements each containing 21
(for Ne) and 23 (for Ar) DVR functions. The absorbing boundary is
implemented by a cos$^{1/4}$ mask function switched on at $r = 180$ to avoid
reflection from the box boundary.
{\color{black}
The ground state of each method is obtained by the imaginary
time relaxation, and then the real-time dynamics are simulated starting from
the ground state.
}
The Fourth-order exponential Runge-Kutta method
\cite{exponential_integrator} is used to propagate the EOMs with 10000
time steps for each optical cycle. 

We compare the performance of the following methods:
(i) TDHF, (ii) TD-MCSCF with a TDHF determinant and singly and
doubly excited configurations included (TD-OCISD), (iii) TD-OCCD, (iv) TD-OCCDT, and
(iv) TD-CASSCF as the fully correlated reference for a given number of
active orbitals. 
In all the methods, the 1$s$ orbital of Ne, and 1$s$2$s$2$p$ orbitals of Ar are kept frozen
at the canonical Hartree-Fock orbitals, and 
for the correlated approaches (ii)-(iv) the eight valence electrons
are correlated among 13 active orbitals.
The method (ii) was first introduced in
Ref.~\citenum{miyagi2014time}, and referred to as the {\color{black}time-dependent}
restricted-active-space self-consistent-field method. It can also be
considered a specialization of more general time-dependent
occupation-restricted multiple-active-space method\cite{sato2016time}. 
Here we denote the method as TD-OCISD (time-dependent optimized CI with
singles and doubles) for simplicity. 

\begin{widetext}
 
\begin{figure}[t!]
\centering
\includegraphics[width=1.0\linewidth]{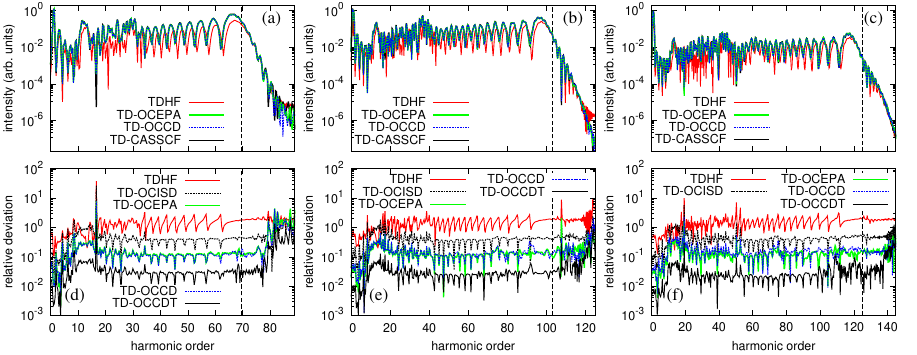}
\caption{\label{fig:nehhg}
The HHG spectra from Ne irradiated by a laser pulse
of a wavelength of 800 nm obtained with various methods (top), and the
relative deviation of the spectral amplitude from the TD-CASSCF spectrum (bottom) defined as Eq.~(\ref{eq:dhhg}), for
laser intensities of 5$\times$10$^{14}$ W/cm$^2$
(a,d), 8$\times$10$^{14}$ W/cm$^2$ (b,e), and 1$\times$10$^{15}$ W/cm$^2$ (c,f). }
\end{figure}
\end{widetext}

\begin{figure}[b!]
\centering
\includegraphics[width=.8\linewidth]{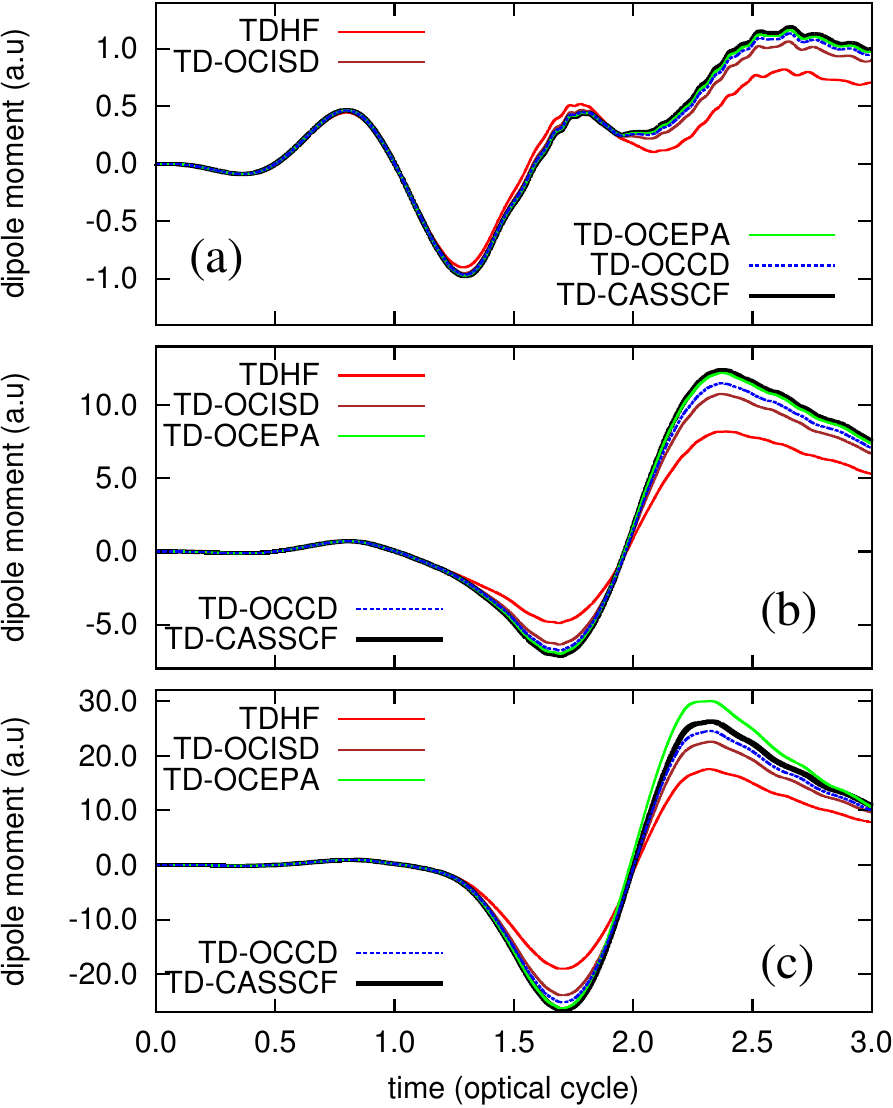}
\caption{\label{fig:ardip}
Time evolution of the dipole moment of Ar irradiated by a laser pulse
of a wavelength of 800 nm and intensities of 2$\times$10$^{14}$ W/cm$^2$
(a), 4$\times$10$^{14}$ W/cm$^2$ (b), and 6$\times$10$^{14}$ W/cm$^2$ (c),
calculated with TDHF, TD-OCISD, TD-OCEPA, TD-OCCD, and TD-CASSCF methods.} 
\end{figure}

\begin{figure}[b!]
\centering
\includegraphics[width=.8\linewidth]{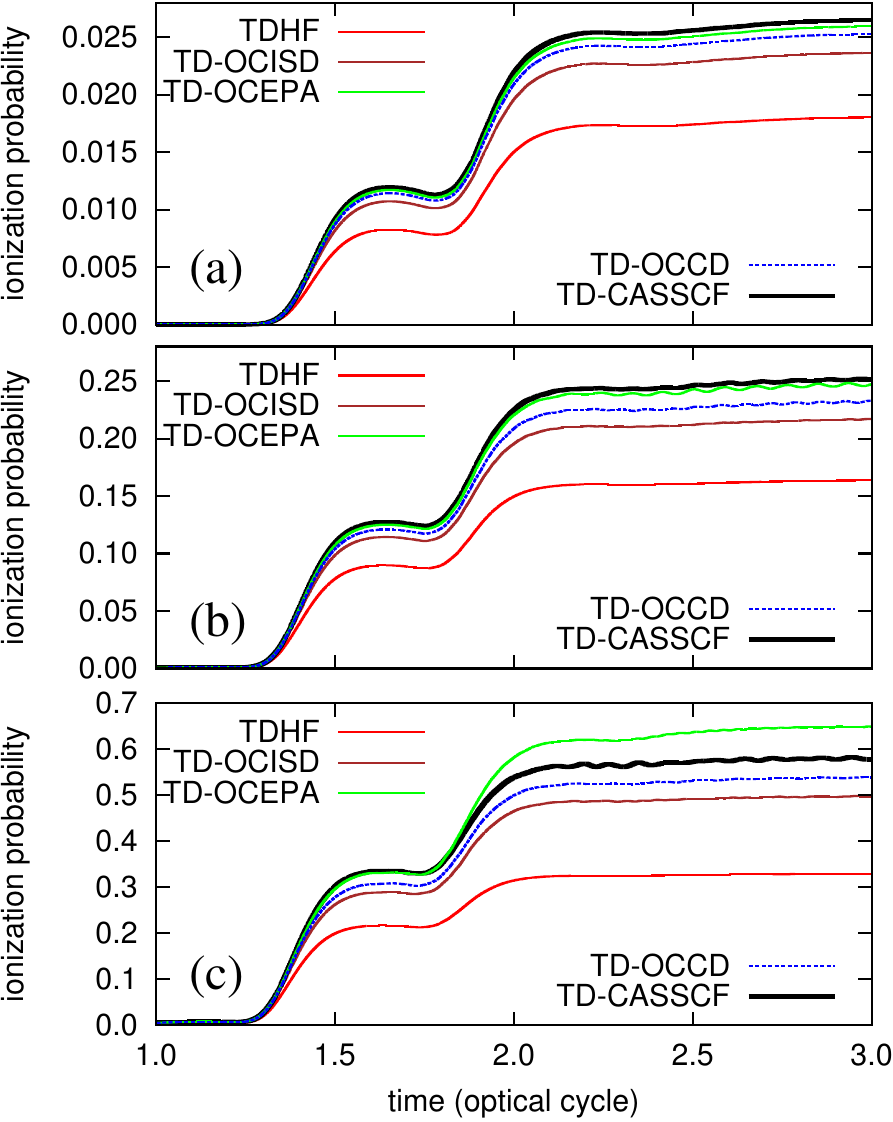}
\caption{\label{fig:arion}
Time evolution of the single ionization
 probability of Ar irradiated by a laser pulse
of a wavelength of 800 nm and intensities of 2$\times$10$^{14}$ W/cm$^2$
(a), 4$\times$10$^{14}$ W/cm$^2$ (b), and 6$\times$10$^{14}$ W/cm$^2$ (c),
calculated with TDHF, TD-OCISD, TD-OCEPA, TD-OCCD, and TD-CASSCF methods.} 
\end{figure}

\begin{widetext}

\begin{figure}[t!]
\centering
\includegraphics[width=1.0\linewidth]{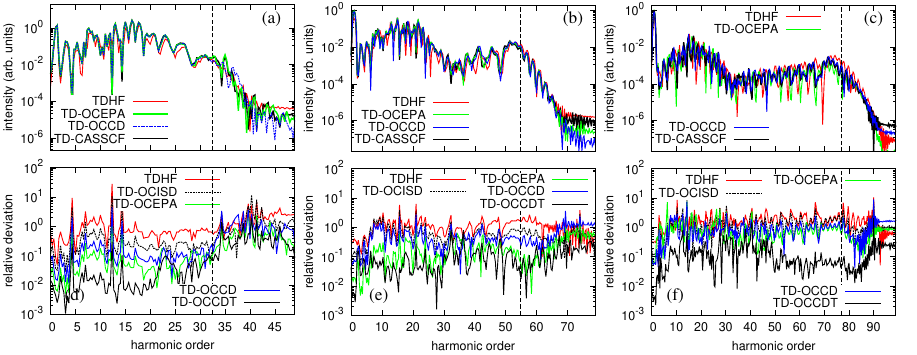}
\caption{\label{fig:arhhg}
The HHG spectra from Ar irradiated by a laser pulse
of a wavelength of 800 nm obtained with various methods (top), and the relative deviation
of the spectral amplitude from the TD-CASSCF spectrum (bottom) defined as Eq.~(\ref{eq:dhhg}), for
laser intensities of 2$\times$10$^{14}$ W/cm$^2$
(a,d), 4$\times$10$^{14}$ W/cm$^2$ (b,e), and 6$\times$10$^{14}$ W/cm$^2$ (c,f) }
\end{figure}
\end{widetext}

The simulation results are given in
Figs.~\ref{fig:nedip}-\ref{fig:nehhg} for Ne and
Figs.~\ref{fig:ardip}-\ref{fig:arhhg} for Ar. For each atom, we report
the time evolution of the dipole moment [Figs.~\ref{fig:nedip} and \ref{fig:ardip}], the
ionization probability [Figs.~\ref{fig:neion} and \ref{fig:arion}], and
the HHG spectra [Figs.~\ref{fig:nehhg} and \ref{fig:arhhg}]. 
The dipole moment is evaluated as a trace $\langle \psi_p|\hat z
|\psi_q\rangle \rho^q_p$ using the
1RDM, and the ionization probability is
defined as the probability of finding an electron outside a sphere of
radius 20 a.u, computed by an expression using {\color{black}1RDM and} 2RDM\cite{Lackner:2015}. 
The HHG spectrum is obtained as the modulus squared $I(\omega)=|a(\omega)|^2$ of the Fourier
transform of the expectation value of the dipole acceleration, 
which, in turn, is obtained with a modified Ehrenfest expression\cite{sato2016time}. 
The TD-OCCDT results of the dipole moment and the ionization probability
are not shown, since they meet a virtually perfect agreement with
TD-CASSCF ones. 
The HHG spectrum is shown only for 
TDHF, TD-OCEPA0, TD-OCCD, and TD-CASSCF methods
[Figs.~\ref{fig:nehhg}(a)-(c) and \ref{fig:arhhg}(a)-(c)] to
keep a visibility, and an absolute relative deviation
\begin{eqnarray}\label{eq:dhhg}
\delta(\omega) = 
\left|
\frac{a(\omega)-a_\textrm{TD-CASSCF}(\omega)}{a_\textrm{TD-CASSCF}(\omega)}
\right|
\end{eqnarray}
of the spectral amplitude $a(\omega)$ from the TD-CASSCF value is reported for each method
[Figs.~\ref{fig:nehhg}(d)-(f) and \ref{fig:arhhg}(d)-(f)].
{\color{black}
Note that both $I(\omega)$ and $\delta(\omega)$ are plotted in the logarithmic scale,
and $\delta(\omega)=1$ corresponds to a 100\% deviation from the
TD-CASSCF amplitude. 
}

The dipole moment (Fig.~\ref{fig:nedip}) and the ionization
probability (Fig.~\ref{fig:neion}) of Ne atom show a general trend 
that the deviation of results for each method from TD-CASSCF ones decreases as
TDHF $\gg$ TD-OCISD $>$ TD-OCEPA0 $\approx$ TD-OCCD $>$ TD-OCCDT.
The HHG spectra [Fig.~\ref{fig:nehhg}~(a)-(c)] are well reproduced by
all the methods, except a systematic underestimation of the intensity
by TDHF, with the magnitude of the error amounting to 100\% of the
TD-CASSCF spectral amplitude as shown in Fig.~\ref{fig:nehhg}~(d)-(f). 
The magnitude of the error depends weakly on the harmonic order at the
plateau region, which decreases again as TDHF $\gg$ TD-OCISD $>$
TD-OCEPA0 $\approx$ TD-OCCD $>$ TD-OCCDT. 
The Ne atom is characterized by its large ionization potential of 21.6
eV, resulting in relatively low ionization probabilities
(Fig.~\ref{fig:neion}) for the present laser pulses. In that case,
TD-OCEPA0 and TD-OCCD give a notably similar, and accurate description of dynamics,
implying that the truncation after the doubles amplitudes and the
linearization of the Lagrangian [Eq.~\ref{eq:L_CEPA0}] are both justified.

The Ar atom, having a lower ionization potential of 15.6 eV,
exhibits a richer dynamics than Ne, e.g, 
a larger-amplitude oscillation of the dipole moment
(Fig.~\ref{fig:ardip}), ionization probabilities
as high as 70\% (Fig.~\ref{fig:arion}), and
HHG spectra characterized by a dip around 34th harmonic related to the
Cooper minimum of the photoionization spectrum\cite{Worner:2009} at the same energy
(Fig.~\ref{fig:arhhg}). The trend of the accuracy of each method is similar to the
case for Ne, except that TD-OCEPA0 and TD-OCCD 
give noticeably different descriptions
of dynamics.
In general, TD-OCEPA0 tends to capture a larger part of the
correlation effect (the difference between TDHF and TD-CASSCF) than
TD-OCCD does.
This results in, on one hand, a better agreement of the TD-OCEPA0
results (than TD-OCCD) with TD-CASSCF ones for lower
intensity cases [Fig.~\ref{fig:ardip}~(a),(b),
Fig.~\ref{fig:arion}~(a),(b), and Fig.~\ref{fig:arhhg}~(d),(e)], and, on the other, leads to an overestimation of the
correlation effect for the highest intensity [Fig.~\ref{fig:ardip}~(c),
Fig.~\ref{fig:arion}~(c), Fig.~\ref{fig:arhhg}~(f)], somewhat analogous
to the overestimation of the ground-state correlation energy as
discussed in Sec.~\ref{sec3-1}.

The nonlinear exponential parametrization in TD-OCCD seems to play a
role in correcting the overestimation, and the inclusion of the triple
excitations (TD-OCCDT) is essential to retain the decisive accuracy
across the employed range of the laser intensity.
Despite the aforementioned overestimation of the correlation effect for
higher intensities,  
we judge that the present results of TD-OCEPA0 are quite encouraging; it clearly
outperforms TD-OCISD for all properties of both atoms, and at least 
performs equally as TD-OCCD for a moderate laser intensity.

Finally, we compare the computational cost of TD-OCEPA0 and TD-OCCD
methods.
Table~\ref{tab:timing_electron} reports the central processor unit (CPU) time for
the computational bottlenecks differing in these
two methods, with varying numbers of active electrons and active
orbitals. The largest active space with 18 electrons and 18 orbitals
(18e-18o) is challenging for the fully correlated TD-CASSCF.

\begin{table}[!t] 
\caption{\label{tab:timing_electron} Comparison of the CPU
time (in second) spent for the evaluation of the $T_2$ equation,
$\Lambda_2$ equation, and 2RDM for TD-OCCD and
TD-OCEPA0 methods with various active spaces. See text for more details.}
\begin{ruledtabular}
\begin{center}
\begin{tabular}{crrrrrrr}
& \multicolumn{3}{c}{TD-OCCD} && \multicolumn{3}{c}{TD-OCEPA0} \\
\cline{2-4} \cline{6-8}
\multicolumn{1}{c}{active space\footnotemark[1]}
&\multicolumn{1}{c}{T$_2$}&\multicolumn{1}{c}{$\Lambda_2$}&\multicolumn{1}{c}{2RDM}&
&\multicolumn{1}{c}{T$_2$}&\multicolumn{1}{c}{$\Lambda_2$}&\multicolumn{1}{c}{2RDM} \\
\hline \\
(8e{\rm-}9o)  &\,  8.1&\, 11.4&\,  20.7&\,&\,  3.3&\,{-}&\,  3.1 \\
(8e{\rm-}13o) &\, 40.8&\, 55.5&\, 109.4&\,&\, 18.2&\,{-}&\, 19.8 \\
(8e{\rm-}20o) &\,254.9&\,332.1&\, 703.9&\,&\,131.4&\,{-}&\,187.9 \\
(14e{\rm-}16o)&\,248.2&\,307.2&\, 555.5&\,&\,111.1&\,{-}&\, 83.2 \\
(16e{\rm-}17o)&\,314.4&\,437.0&\, 852.1&\,&\,131.5&\,{-}&\,124.4 \\
(18e{\rm-}18o)&\,452.6&\,619.6&\,1024.8&\,&\,187.9&\,{-}&\,143.3 \\
\end{tabular}
\end{center}
\end{ruledtabular}
\footnotetext[1]
 {CPU time spent for the simulation of Ar atom with $n_{\rm act}$ active orbitals and
 $N_{\rm act}$ active electrons ($n_{\rm act}$e-$N_{\rm act}$o), 
 recorded and accumulated over 1000 time steps of a real-time simulation
 ($I_0=2\times 10^{14}$ W/cm$^{2}$ and $\lambda=800$ nm.), 
 using {\color{black}an} Intel(R) Xeon(R) CPU with 12 processors having a clock speed of {\color{black}3.33GHz}.}
\end{table}


First, depending on the active space
configurations, the evaluation of the $\hat{T}_2$ equation of the
TD-OCEPA0 [Eq.~(\ref{eq:td-ocepa0_t2})] is 1.9$\sim$2.5 times faster 
than that of TD-OCCD [Eq.~(\ref{eq:td-occd_t2})]. 
A bigger computational gain comes from the fact that one need not solve for
the $\Lambda_2$ equation, which, for TD-OCCD
[Eq.~(\ref{eq:td-occd_l2})] takes {\color{black}longer} than that 
for the $\hat{T}_2$ equation because of more mathematical operations
involved.
A further significant cost reduction is obtained by TD-OCEPA0 for the 2RDM
evaluation [Eqs.~(\ref{eq:td-ocepa0_den2})], which is 5.5$\sim$7.2 times
faster than the TD-OCCD case [Eqs.~(\ref{eq:td-occd_den2})].  
As a whole, the TD-OCEPA0 simulation with, e.g, the (18e-18o) active space
achieves 6.3 times speed up relative to the TD-OCCD simulation with
the same active orbital space.

\section{Summary\label{sec4}}
We have presented the implementation of TD-OCEPA0 method as a
cost-effective approximation within the TD-OCC framework, for the first 
principles study of intense laser-driven multielectron dynamics.
The TD-OCEPA0 method retains the important size-extensivity and
gauge-invariance of TD-OCC, and computationally much more
efficient than the full TD-OCCD method. 
As a first numerical test, we applied the
{\color{black}present implementation}
to Ne and Ar atoms irradiated by an intense near infrared laser pulses
with three different intensities to compare the time-dependent dipole moment,
the ionization probability, and HHG spectra with those obtained with
other methods including the fully correlated TD-CASSCF methods with the
same number of active orbitals.
It is observed that, for the highest laser intensity, with sizable
ionization, the TD-OCEPA0 tends to
overestimate the correlation effect defined as the difference between
TDHF and TD-CASSCF descriptions. For moderate intensities, however, the
TD-OCEPA0 method performs at least equally well as TD-OCCD with a
substantially lower computational cost. It is anticipated that
the present TD-OCEPA0 method serves as an important theoretical tool to
investigate ultrafast and/or high-field processes in chemically relevant
large molecular systems. 

\begin{acknowledgments}
This research was supported in part by a Grant-in-Aid for
Scientific Research (Grants No. 16H03881, No. 17K05070,
No. 18H03891, and No. 19H00869) from the Ministry of Education, Culture,
Sports, Science and Technology (MEXT) of Japan. 
This research was also partially supported by JST COI (Grant No.~JPMJCE1313), JST CREST (Grant No.~JPMJCR15N1), and by MEXT Quantum Leap Flagship Program (MEXT Q-LEAP) Grant Number JPMXS0118067246.
\end{acknowledgments}

\appendix
\section{Algebraic details of TD-OCCD}\label{app:td-occd}
The TD-OCCD method implemented in Ref.~\citenum{sato2018communication}
employs the same truncation ($\hat{T}=\hat{T}_2$ and
$\hat{\Lambda}=\hat{\Lambda}_2$) as for TD-OCEPA0, but
retains the full exponential operator $e^{\hat{T}_2}$. As a result, the amplitude
EOMs are given by
%
\begin{eqnarray}
i\dot{\tau}^{ab}_{ij}
&=&\label{eq:td-occd_t2}
v_{ij}^{ab}-p(ij) \bar{f}_j^k\tau_{ik}^{ab}+p(ab) \bar{f}_c^a \tau_{ij}^{cb} \nonumber \\
&+&\frac{1}{2} v_{cd}^{ab}\tau_{ij}^{cd}
+\frac{1}{2} v_{ij}^{kl} \tau_{kl}^{ab}+p(ij)p(ab)
v_{ic}^{ak} \tau_{kj}^{cb} \nonumber \\ 
&-&\frac{1}{2}p(ij) \tau_{ik}^{ab} \tau_{jl}^{cd} v_{cd}^{kl}
+\frac{1}{2}p(ab) \tau_{ij}^{bc} \tau_{kl}^{ad} v_{cd}^{kl} \nonumber \\
&+&\frac{1}{4} \tau_{kl}^{ab} \tau_{ij}^{cd} v_{cd}^{kl}
+\frac{1}{2}p(ij)p(ab) \tau_{il}^{bc} \tau_{jk}^{ad} v_{cd}^{kl},
\end{eqnarray}
\begin{eqnarray}
-i\dot{\lambda}^{ij}_{ab}
&=&\label{eq:td-occd_l2}
v_{ab}^{ij}-p(ij) \bar{f}_k^i
\lambda_{ab}^{kj}+p(ab) \bar{f}_a^c\lambda_{cb}^{ij} \nonumber \\
&+&\frac{1}{2} v_{ab}^{cd}\lambda_{cd}^{ij}
+\frac{1}{2} v_{kl}^{ij}\lambda_{ab}^{kl}+p(ij)p(ab)
v_{kb}^{cj}\lambda_{ac}^{ik} \nonumber \\ 
&-&\frac{1}{2}p(ij) \lambda_{cd}^{ik} \tau^{cd}_{kl} v_{ab}^{jl}
+\frac{1}{2}p(ab) \lambda_{bc}^{kl} \tau^{cd}_{kl} v_{ad}^{ij} \nonumber \\
&+&\frac{1}{4} \lambda_{ab}^{kl} \tau_{kl}^{cd} v_{cd}^{ij}
+\frac{1}{2}p(ij)p(ab) \lambda_{ac}^{jk} \tau_{kl}^{cd} v_{bd}^{il} \nonumber \\
&-&\frac{1}{2}p(ij) \lambda_{ab}^{ik} \tau_{kl}^{cd} v_{cd}^{jl}
+\frac{1}{2}p(ab) \lambda_{bc}^{ij} \tau_{kl}^{cd} v_{ad}^{kl} \nonumber \\
&+&\frac{1}{4} \lambda_{cd}^{ij} \tau_{kl}^{cd} v_{ab}^{kl}. 
\end{eqnarray}

The EOMs for orbitals are
formally the same as that for TD-OCEPA0,
Eqs.~(\ref{eq:eom_orb})-(\ref{eq:eom_orb_p}),
but with RDMs $\rho$ replaced with Hermitialized ones, 
$D^p_q=(\rho^p_q+\rho^{q*}_p)/2$ and
$P^{pr}_{qs}=(\rho^{pr}_{qs}+\rho^{qs*}_{pr})/2$.
Finally the algebraic expression for non-zero correlation RDM elements is
\begin{subequations}
\begin{eqnarray}
\gamma^{ j }_{ i }&=&\label{eqs:td-occd_den1}
- \frac{ 1 }{ 2 }
\lambda^{ k j }_{ c d } 
\tau^{ c d }_{ k i },
\gamma^{ b }_{ a }=
\frac{ 1 }{ 2 }
\lambda^{ k l }_{ c a } 
\tau^{ c b }_{ k l }, \\
\gamma^{ c d }_{ a b }&=&
\frac{ 1 }{ 2 }
\lambda^{ k l }_{ a b } 
\tau^{ c d }_{ k l },
\gamma^{ k l }_{ i j }=
\frac{ 1 }{ 2 }
\lambda^{ k l }_{ c d } 
\tau^{ c d }_{ i j }, \\
\gamma^{ i a }_{ b j }&=&
\lambda^{ k i }_{ c b }
\tau^{ c a }_{ k j },
\gamma^{ i j }_{ a b }=
\lambda^{ i j }_{ a b }, \\
\gamma^{ a b }_{ i j }&=&\label{eq:td-occd_den2}
\tau^{ a b }_{ i j }
+\frac{ 1 }{ 2 }
p(ij)p(ab)
\lambda^{ k l }_{ c d }
\tau^{ c a }_{ k i }
\tau^{ b d }_{ j l } \nonumber \\
&-&
\frac{ 1 }{ 2 }
p( i j )
\lambda^{ k l }_{ c d }
\tau^{ c d }_{ k i }
\tau^{ a b }_{ l j }
-\frac{ 1 }{ 2 }
p( a b )
\lambda^{ k l }_{ c d }
\tau^{ c a }_{ k l }
\tau^{ d b }_{ i j } \nonumber \\
&+&
\frac{ 1 }{ 4 }
\lambda^{ k l }_{ c d }
\tau^{ c d }_{ i j }
\tau^{ a b }_{ k l }.
\end{eqnarray}
\end{subequations}

\end{document}